\documentclass{article}
\usepackage{emulateapj}
\usepackage{amsmath}
\usepackage{multirow}
\usepackage{epsf}
\usepackage{rotating}
\usepackage{lscape}
\usepackage{flushrt}

\def\apjl{{\it Ap. J. (Lett.)}}

\def\arcmin{$^\prime$}

\def\h0{$H_{0}=50$~km~sec$^{-1}$~Mpc$^{-1}$}
\def\q0{$q_{0}$}

\def\etal    {{et~al.}~}

\def\be{\begin{equation}}
\def\ee{\end{equation}}
\def\bea{\begin{eqnarray}}
\def\eea{\end{eqnarray}}

\lefthead{Donnelly et al.}
\righthead{Merging Binary Clusters}

\slugcomment{accepted for publication in {\em The Astrophysical Journal}}
\begin{document}

\title{Merging Binary Clusters}

\author{R.~Hank~Donnelly\altaffilmark{1}, W.~Forman\altaffilmark{1}, C.~Jones\altaffilmark{1}, H.~Quintana\altaffilmark{2},
A.~Ramirez\altaffilmark{3}, E.~Churazov\altaffilmark{4,}\altaffilmark{5}, M.~Gilfanov\altaffilmark{4,}\altaffilmark{5}}

\altaffiltext{1}{Harvard-Smithsonian Center for Astrophysics, 60 Garden Street,
Cambridge, MA 02138, USA} 
\altaffiltext{2}{Department of Astronomy and Astrophysics, Pontificia
Universidad Catolica de Chile, Casilla 104, Santiago 22, CHILE}
\altaffiltext{3}{Department of Physics, Universidad de La Serena, Benavente 
980, La Serena, CHILE}
\altaffiltext{4}{Max Planck Institute f\"{u}r Astrophysik, 85740 Garching bei M\"{u}nchen, GERMANY}
\altaffiltext{5}{Space Research Institute (IKI), Moscow 117810, RUSSIA}

\centerline{\date}
\begin{abstract}

We study three prominent bi-modal X-ray clusters: A3528, A1750 and
A3395. Using observations taken with {\em ROSAT} and {\em ASCA}, we
analyze the temperature and surface brightness distributions. We also
analyze the velocity distributions of the three clusters using new
measurements supplemented with previously published data. We examine
both the overall cluster properties, as well as the two sub-cluster
elements in each. These results are then applied to the determination
of the overall cluster masses, and demonstrate excellent consistency
between the various methods used. While the characteristic parameters
of the sub-clusters are typical of isolated objects, our temperature
results for the regions between the two sub-clusters clearly confirm
the presence of merger activity. These three clusters represent a
progression of equal-sized sub-cluster mergers, starting from initial
contact to immediately before first core passage.

\end{abstract}


\section{INTRODUCTION}
Optical and X-ray studies (e.g.  Forman \etal 1981; Geller \& Beers
1982; Jones \& Forman 1984; Dressler \& Shectman 1988; Mohr et
al. 1995; Bird 1994; Slezak et al. 1994) have shown that galaxy
clusters are dynamically evolving systems, exhibiting a variety of
substructure and asymmetric morphologies. The high frequency of
substructure-- $\sim 40$\% (Forman \& Jones 1990)-- suggests that
clusters are still forming hierarchically through the merger and
accretion of subclusters and galaxy groups. While the prevalence of
substructure has primarily been used to constrain cosmologies
(Richstone, Loeb, \& Turner 1992; Mohr et al. 1995), detailed studies
of merging clusters impact a number of important areas including
cooling flow formation and evolution, galaxy evolution, and
gravitational mass measurements.

Temperature maps of the X-ray emitting gas are an especially sensitive
tool for the detection of dynamic activity. Hydrodynamic simulations
indicate that mergers should produce characteristic temperature
patterns that survive 4-6 times longer than perturbations in the gas
density (e.g.  Schindler \& M\"uller 1993; Ricker 1997). Recent
observations (e.g. Henry \& Briel 1995; Henriksen \& Markevitch 1996;
Markevitch \etal 1996a, 1998; Donnelly \etal 1998, Markevitch et
al. 2000) have found just such characteristic temperature structures
in a variety of clusters. In particular, in the earliest stages of a
merger, simulations show the development of a shock located between
the two subclusters and significant heating of the local ICM (Evrard
1990a and b; Schindler \& M\"{u}ller 1993).

Abell 3528, Abell 1750 and Abell 3395 are known as canonical binary
galaxy clusters (Forman et al. 1981, Raychaudury et al. 1991,
Henriksen and Jones 1996). All three exhibit two clearly separated
peaks of emission in X-rays, and varying degrees of distortion in
their X-ray surface brightness suggesting a progressive sequence of
merging.

We used X-ray data from the {\em ROSAT} and {\em ASCA} satellites to
characterize the emission profile of the gas as well as to map the
distribution of temperatures and fit specific regions of interest with
typical spectral models. We also analyzed optical velocity
measurements to study the internal kinematics and overall dynamics of
each cluster.

In Section~\ref{sec:obs} we discuss the data, its reduction and
present the basic results of our work.  Section~\ref{sec:results}
gives estimates of the masses determined from fitting the emission
intensity profile, the X-ray luminosity and a virial estimate based on
the velocities of the galaxies. Finally, we discuss the implications
of our results in Sections~\ref{sec:discussion} and \ref{sec:summary}
respectively.  Throughout the paper we assumed $\rm H_0 = 65\ km\
s^{-1}\ Mpc^{-1}$, and the error bars quoted on all results are at the
1 $\sigma$ level.

\section{OBSERVATIONS \& METHODS}\label{sec:obs}
All three clusters-- A3528, A1750, and A3395-- were observed using
both {\em ROSAT} and {\em ASCA}. Data were obtained from the HEASARC
public archives and the details of the observations are given in
Table~\ref{tab:obsdat}.  Using the broad energy response of {\em ASCA}
(0.5-10.0 keV), these observations were used to accurately measure the
gas temperatures and distributions in these clusters. The
significantly better spatial resolution of {\em ROSAT} was used to
determine the cluster surface brightness distribution, while {\em
ROSAT} spectra of the central regions provided a diagnostic for the
presence of cool gas.

\placetable{tab:obsdat}

We also compiled a list of velocities and positions of galaxies
belonging to the three clusters using new data, supplemented with
previously measured redshifts available from the literature. Using
this data we analyzed the mean velocities and dispersions of the three
clusters as well as sub-samples selected based on the X-ray
emission. These results were then used to estimate the virial masses
of the merging sub-clusters as well as a comparison to the X-ray
results.

\subsection{{\em ASCA} GIS}
The {\em ASCA} GIS observations were ``cleaned'' using standard
processing tools (Arnaud 1993) with a conservative version of the
filtering criteria (ABC Guide\footnote{See
http://heasarc.gsfc.nasa.gov/docs/asca/abc/abc.html }).  The data from
both GIS detectors was combined in our final analysis, while the SIS
detectors, with their smaller field of view, were not used for this
study.

To correctly characterize the temperature distribution of these
extended sources, a correction for the energy dependent PSF must be
applied (Takahashi et al. 1995).  We employed a method developed by
Churazov and Gilfanov (Churazov et al. 1996, hereafter the CG method),
and used previously on other clusters (Donnelly et al. 1998, Donnelly
et al. 1999, Churazov et al. 1999, and Henricksen et al. 2000).  This
method first explicitly corrects the core ($r\le 6'$) PSF, then
approximates the PSF of the wings via a Monte-Carlo simulation. The
temperature is then fit in each $15''$ pixel using a linear
combination of two fiducial single temperature spectra and smoothed at
the resolution scale of the telescope ($\sim 5'$) to reduce noise.  As
shown for A1367 (Donnelly et al. 1998) the results of this method are
fully consistent with those found using the method presented by
Markevitch et al. (1996 and 1998).

\placetable{tab:struc+T}

Applying this method to each of our three clusters, we fit a
Raymond-Smith model to the emission from the entire cluster to derive
a global gas temperature ($T_{ASCA}$).  The results are given in
Table~\ref{tab:struc+T}.  However, the CG method also allowed us to
generate two dimensional continuous maps of the gas temperature
distributions within the clusters. These temperature maps are shown in
Figures~\ref{fig:3528}-\ref{fig:3395} (top left).

\placefigure{fig:3528}

\placefigure{fig:1750}

\placefigure{fig:3395}

Finally, we also used the CG method to define discrete regions on each
cluster and fit the temperature within each region, as shown in
Figures~\ref{fig:3528}-\ref{fig:3395}.  In each cluster we defined a
region around the core of each sub-cluster, two regions to the
``outside''-- i.e. away from the other sub-component, and a region
between the two subclusters.  The fit temperatures, with the
associated uncertainties, for these regions are shown in the bottom
left of Figures~\ref{fig:3528}-\ref{fig:3395}.

\subsection{{\em ROSAT} PSPC}
The {\em ROSAT} PSPC data were reduced using the standard procedures
outlined by Snowden (1994; see also Snowden et al. 1994). By combining
only the data from bands 4 through 7 (0.44-2.04 keV), we excluded the
lowest energies, that generally have higher X-ray background.

For each of the six sub-clusters in the sample, we generated radial
surface brightness profiles in 1\arcmin~ annuli centered on the peak
emission for each sub-component. The annuli extended from 0\arcmin~ to
45\arcmin~ and, to minimize contamination of the intensity profile, we
excluded the azimuthal half of the profile toward each sub-component's
partner. We then measured the average surface brightness ($\rm cts\
s^{-1}\ arcmin^{-2}$) in each annulus and fit the resultant surface
brightness profile with a standard hydrostatic, isothermal
$\beta$-model:
\begin{equation}
\Sigma(r)=\Sigma_0\left[1+\left(\frac{r}{r_c}\right)^2\right]^{-(3\beta-\frac{1}{2})}
\end{equation}
(Cavaliere \& Fusco-Femiano 1976).  The backgrounds were fit by
including a constant component in our models. For A1750 and A3395 the
backgrounds fell in the range $\rm 2.08-2.80\times 10^{-4}\ cts\
s^{-1}\ arcmin^{-2}$, consistent with typical PSPC backgrounds. For
both clusters the errors for the two sub-clusters were consistent with
single constant backgrounds. A3528 lies near the edge of its {\em
ROSAT} frame (an AGN is actually the target of this observation). We
used a nearby region of sky to determine the background and found a
count rate of $\rm 2.43\times 10^{-4}\ cts\ s^{-1}\ arcmin^{-2}$ that
was included in our fits for this cluster. The bolometric luminosity
for each sub-cluster was also calculated using the same regions.  The
global fit results and luminosities are given in
Table~\ref{tab:struc+T}.

To search for evidence of the presence of cool gas within the cores,
we also extracted {\em ROSAT} spectra from the central 2\arcmin~ of
each sub-cluster. Using XSPEC to fit a single temperature
Raymond-Smith model to the data, we confirmed the presence of cool gas
in the NE sub-cluster of A1750 ($T=1.9^{+1.3}_{-0.5}$ keV) and both
subclusters in A3528 ($T_{SE}=2.1^{+0.7}_{-0.4}$ keV and
$T_{NW}=2.3^{+1.3}_{-0.5}$ keV).

\subsection{OPTICAL SPECTROSCOPY}
A3395 was observed with the ARGUS fiber spectrograph at the CTIO 4m
telescope, and with the Shectograph detector at the $100''$ DuPont
telescope at the Las Campanas Observatory(LCO). The
Shectograph/$100''$ combination at LCO also was used for all of the
spectroscopic observations of A1750.  Specifics of the instrumental
set ups, observing details, reduction and calibration procedures, as
well as an evaluation and discussion of internal and external errors,
are given in Quintana \& Ram\'{i}rez (1990). For A3528 the data were
drawn from the work by Quintana et al. (2000) on the Shapley
Supercluster. The interested reader is directed to that work for a
complete discussion of the instrumental configuration.

Velocity determinations were carried out using both a
cross-correlation technique and by identifying and fitting line
profiles. All reductions were performed using the LONGSLIT, ONEDSPEC
and RVSAO packages in IRAF. No systematic internal errors or zero
point corrections were found. Typical uncertainties for individual
spectra were 30-40 km s$^{-1}$ using the line fitting
method. Cross-correlation was applied following the procedures
described in Quintana, Ram\'{\i}rez, \& Way 1995 (QRW95) using 8
templates. In this case the errors were somewhat higher (40-50 km
s$^{-1}$) compared to the line fitting method. There was no velocity
offset between the methods, and the cross-correlation was not applied
to spectra with emission lines.

We expect some systematic shifts between velocities for A3395 derived
from the two different instrument/telescope configurations (LCO 2.5m
and CTIO 4m).  We confirmed this on a wider data sample, as discussed
in QRW95.  Following the discussion there, for the present set, we
combined velocities after correcting for a 50 km s$^{-1}$ shift
between ARGUS and Shectograph data.

For all three clusters we excluded a small number of objects with
obviously discrepant velocities (i.e. a velocity more than 5,000 $\rm
km\ s^{-1}$ greater or lesser than the peak of the cluster
distribution). In the case of A3528 this included 18 objects centered
around 23,000 $\rm km\ s^{-1}$.  These objects are scattered around
the field and may represent a background group.

The data for all three clusters also were supplemented by redshifts
available through the NASA/IPAC Extragalactic Database (NED). For
A3528 four velocities were drawn from Katgert et al. (1998), although,
no attempt at a determination of a systematic offset was attempted due
to the small sample size. The full listing is given in
Table~\ref{tab:vel1}.

\placetable{tab:vel1}

The NED data for A1750 are from Beers et al. (1991) which shares 26
galaxies with our new data (67 galaxies).  From a comparison of the
velocities of these common objects, we find that our velocities are
systematically larger by 69 km s$^{-1}$.  In order to have all of our
data in a self-consistent frame, we have applied this offset to the
NED data and the offset values are listed in Table~\ref{tab:vel2}.

\placetable{tab:vel2}

Similarly for A3395, we have 54 Shectograph velocities in common with
Teague, Carter \& Gray (1990, TCG90) with a systematic offset of $\rm
82\ km s^{-1}$.  In the case of the ARGUS data, there are 24
velocities in common with TCG90 and our velocities are larger by $\rm
133\ km s^{-1}$. Both offsets were applied to the NED data and final
velocities are listed in Table~\ref{tab:vel3}. There are two galaxies
observed with ARGUS and published in TCG90 with large discrepant
velocities, both with a TCG90 R value below $\sim$2.5, that we
included in the cluster redshift, contrary to TCG90 who believe them
to be background galaxies.

\placetable{tab:vel3}

When galaxies had more than one velocity, we adopted as the final
value the weighted mean velocity and the associated uncertainty.
Weighting factors came from the published uncertainties combined with
our internal uncertainties. More details about this procedure are
extensively discussed in QRW95.

Using the statistical prescriptions described in Beers et al.  (1990)
we determined the ``mean'' velocity for each of our three clusters
(see Tables~\ref{tab:struc+T} and ~\ref{tab:compvel}).  We find
generally excellent agreement with previous results (Abell, Corwin, \&
Olowin 1989, Struble \& Rood 1999). One exception is that for
A3528. Our results give a ``mean'' velocity more than 500 $\rm km\
s^{-1}$ larger than that found by Struble \& Rood (1999), but which is
more consistent with the values found previously for this cluster by
Abell, Corwin, \& Olowin (1989) and Katgert et al. (1996).

\placetable{tab:compvel}

The positions of the galaxies are shown with the X-ray intensity
isophotes from {\em ROSAT} overlaid in
Figures~\ref{fig:3528}-\ref{fig:3395}. We expanded the field of view
from that shown for the temperature maps to include all of the
galaxies that appear to be members of each cluster. Galaxies with
lower velocities than our ``mean'' cluster velocity are shown in blue
while those with higher velocities are shown in red. A velocity
histogram for each cluster is also presented.

We note that A3395 has a high velocity bump in its distribution. These
objects are shown in Figure~\ref{fig:3395} as open circles. While they
are localized in velocity space, they are distributed very widely
across the field of view. This suggests that it is unlikely that they
are a background group, and thus they have simply been included in
the overall sample.

\section{Results and Analysis}\label{sec:results}

\subsection{MASSES}
We first estimated the mass of the X-ray emitting gas by using the
luminosities for the sub-clusters determined from the {\em ROSAT} data
and solving for the central density (given in Table~\ref{tab:struc+T}),
\begin{eqnarray}
\lefteqn{L(r)=\frac{2\pi n_en_H\Lambda_0r_c^3}{(1-3\beta)}\times}\nonumber\\
&\int_{0}^{\infty}\left[\left(1+s^2+\left[\frac{r}{r_c}\right]^2\right)^{-3\beta+1}-(1+s)^{-3\beta+1}\right]ds
\end{eqnarray}
(David et al. (1990). The gas density distribution is then integrated
to the radius of interest,
\begin{equation}
M_{gas}(r)=4\pi\rho_0\int_{0}^{r}s^2\left[1+\left(\frac{s}{r_c}\right)^2\right]^{-\frac{3\beta}{2}}ds
\end{equation}
Gas masses within 0.5 and 1.0 Mpc are given in Table~\ref{tab:masses}.

\placetable{tab:masses}


We estimated the total mass of each sub-cluster assuming
spherical symmetry, hydrostatic equilibrium and using our fit values
for the core radii and $\beta$'s. With these assumptions the mass
contained within a radius $r$ is,
\begin{equation}
M(<r)=-\frac{kT(r)}{\mu m_pG}\left(\frac{d \log \rho_g(r)}{d\log
r}+\frac{d \log T(r)}{d\log r}\right)r.
\label{eq:mass}
\end{equation}

Markevitch et al. (1998, 1999) found that the temperature profiles of
galaxy clusters are well approximated by a polytropic equation of the
form,
\begin{equation}
T\propto (1+r^2)^{-\frac{3}{2}\beta(\gamma -1)}
\label{eq:Tprof}
\end{equation}
where $\gamma\simeq 1.24$.  Using this function for the temperature,
Equation~\ref{eq:mass} reduces to
\begin{equation}
M(r)=3.70\times10^{13}M_\odot \frac{0.60}{\mu}{T(r)} r\frac{3 \beta
\gamma r^2}{r_c^2+r^2},
\end{equation}
where $T(r)$-- derived from the fit temperature in the core of each
sub-cluster-- and r are measured in keV and Mpc respectively.
However, recent work (Arnaud \etal 2001) done with data taken from
{\em XMM} suggests that clusters are actually isothermal. If this is
the case then our estimates of the total mass are too high by a factor
of 1.24.  Table~\ref{tab:masses} gives the results for both the
non-isothermal and isothermal total masses. We have also included the
total (non-isothermal) mass within 0.5 Mpc for comparison with the gas
mass at the same radius.

Evrard et al. (1996) suggested that cluster scaling relations also
provide reliable estimates of the mass. A volume with an overdensity
of $\delta_c$ has a total mass of 
\begin{equation}
M(r<r_{\delta_c})=\delta_c\frac{4}{3}\pi\rho_{crit}r_{\delta_c}^3.
\end{equation}
Inverting the scaling relation between radius and
temperature (Mohr et al. 2000),
\begin{equation}
r_{\delta_c}=\frac{(a T)^{\frac{1}{2}}}{(\delta_c
\rho_{crit})^{\frac{1}{2}}} 
\end{equation}
to solve for $\delta_c$, we find that
\begin{equation}
M(r<r_{\delta_c})=\frac{4}{3}\pi aTr_{\delta_c}\label{eq:mass2}
\end{equation}
For comparison we calculate the mass contained within 1 Mpc. Taking
the result from Mohr et al.(2000) that $r_{200}=4h^{-1}_{50}\ Mpc$ for
a 10 keV cluster in our cosmology, Equation~\ref{eq:mass2} then
reduces to
\begin{equation}
M(r<1\ Mpc)=9.30\times 10^{14} \left(\frac{T}{10 keV}\right) M_\odot.
\end{equation}
The masses within 1 Mpc for each subcluster using the fit temperature
from the core can be found in Table~\ref{tab:masses}.  Our results
using the scaling relation are generally consistent with those derived
from the luminosity and intensity profiles.

The cluster velocity distributions also demonstrate typical
dispersions with scales of order 1000 $\rm km\ s^{-1}$.  Again
employing the statistical methods of Beers et al. (1990), we examined
the characteristics of velocity distributions of each core.  In order
to isolate each sub-cluster in our analysis we selected only galaxies
that lay within a projected radius of 0.5 Mpc of each sub-cluster
center.  Table~\ref{tab:compvel} gives the details of the velocity
distributions and is consistent with typical sub-cluster sized
systems.

Two circles indicate the radial extent of the two different subcluster
core elements the surface distribution maps of the galaxies in
Figures~\ref{fig:3528},~\ref{fig:1750} and ~\ref{fig:3395}. The
velocity histograms of the two core samples in each cluster have been
color coded to the circles and shown along with the overall velocity
histograms. For A3395 the two cores are in such close proximity that
the 0.5 Mpc circles overlap and there are nine objects in common. The
measured dispersions reported in Table~\ref{tab:compvel} for the two
sub-samples in A3395 were performed both including and excluding the
disputed objects. These objects were excluded from the sub-clump
velocity histograms with the number of disputed galaxies in each bin
indicated above that bin.

Following the approach outlined by Beers, Geller and Huchra (1982), we
estimated the entire total mass of each sub-cluster using virial
techniques.  Assuming that they are bound and the velocity dispersions
are isotropic, the total mass of each sub-cluster is
\begin{equation}
M_{virial}=\frac{3\pi}{G}\sigma^2_r\left\langle\frac{1}{r_p}\right\rangle^{-1},
\end{equation}
where $\sigma_r$ is the velocity dispersion along the line of sight
and $\left\langle\frac{1}{r_p}\right\rangle^{-1}$ is the harmonic mean
projected separation on the sky.  The masses and values for
$\left\langle\frac{1}{r_p}\right\rangle^{-1}$ are given in
Tables~\ref{tab:masses} and ~\ref{tab:struc+T} respectively.  We note
that while the radial restriction of the galaxy sample should not
affect the determination of the velocity dispersion, it may lead to a
small underestimate of the mean harmonic radius and thus also the
calculated virial masses.  Except for the southeast element of A3528,
all of the results are in good agreement with the determination from
the X-ray data of the integrated total mass within 1 Mpc.

For the southeast clump of A3528, whose optically determined virial
mass is twice the X-ray derived total mass, we note that the velocity
distribution for this sub-clump has a very extended tail with two
galaxies having rather large relative velocities ($\rm \delta V=1069\
and\ 1554 km\ s^{-1}$) compared to the mean sub-clump velocity. If
these two galaxies are excluded from the velocity calculations for the
southeastern clump, the dispersion drops from $\rm 930\ km\ s^{-1}$ to
$\rm 513\ km\ s^{-1}$. While the average projected distance increases
slightly with this exclusion, the overall effect is to reduce the
estimated virial mass by nearly a factor of three, which brings it
into much better agreement with the total mass within 1 Mpc.

Our results for the total cluster masses are in good agreement with a
similar study done by Henriksen and Jones (1996) on A3395 and A3528
using {\em ROSAT} and a more limited set of galaxy velocities. We note
that apparent differences in the X-ray luminosities between the two
analysis are due the choice of presenting a bolometric (this work)
versus the {\em ROSAT} luminosity.

\section{Discussion}\label{sec:discussion}

Our data suggest that these three clusters represent progressive
stages of first time merger events prior to first core passage.
A3528, with its very azimuthally symmetric intensity isophotes appears
to be at the earliest stage, when gas in the outer halos of the
sub-clumps is just beginning to interact. In A1750, the effects of the
merger have begun to distort the intensity isophotes, while A3395 with
its clearly disrupted intensity distribution appears to be nearly at
first core passage.

Other than their merger state, all three clusters appear to be
generally normal. They all have gas mass fractions of $\sim 5$\% at
0.5 and $\sim 8$\% at 1.0 Mpc, and their fit $\beta$ values and core
radii (Table~\ref{tab:struc+T}) are typical for galaxy clusters (Jones
\& Forman 1984, Jones \& Forman 1999, and Vikhlinin et al. 1999).
 
The temperature maps of all three clusters have elevated gas
temperatures in the region between the two emission peaks. For all
three, the deviation from the overall mean temperature is significant
at the 90\% confidence level. We note that the heated gas does not
have a significant impact on the global fit temperature for each
cluster because of its significantly smaller contributions to the
overall intensities.

In A3528 the deviation between the merger region (region 3 for all
three clusters) and the other regions is the least pronounced and
could still be consistent with a uniform temperature throughout the
entire cluster-- excepting the far northeastern side where the
temperature is especially low.

We find a similar result for the temperature data from A1750, although
the temperature of the merger region is inconsistent with the
temperature of the southwestern core. This suggests that a shock
region is developing in the gas, as it is compressed between the two
sub-clusters.

Finally, A3395 clearly demonstrates the compression and concomitant
heating of the gas that would be expected in merger events similar to
those modeled by Roettiger, Burns \& Loken (1996; see also Roettiger,
Loken \& Burns 1997) and Norman \& Bryan (1999).


The spectral analysis of the {\em ROSAT} data in the sub-cluster cores
indicates that cool gas is still present in both core elements in
A3528 and the northeastern clump of A1750 but not in the other cores.
This is consistent with a scenario where the gas within the core is
heated as the merger process proceeds.

\begin{figure*}[t]
\centerline{
\includegraphics[width=7.5in]{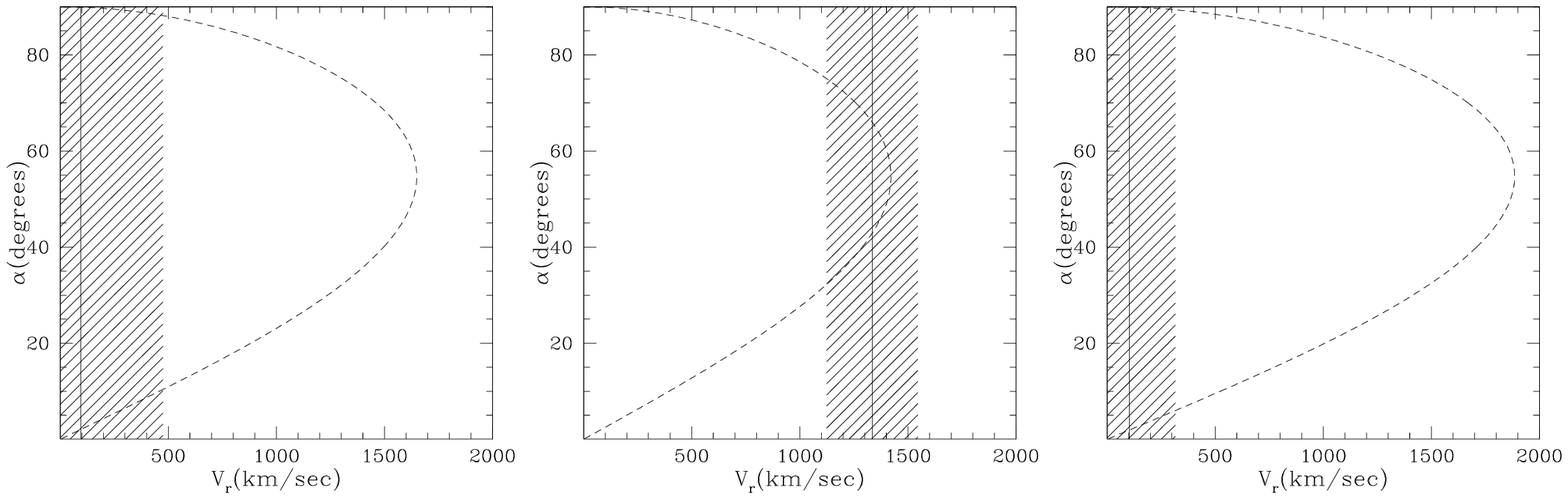}}
\caption{Plots of simple Newtonian energy binding conditions as a
function of measured relative velocity ($V_r$) and projection angle
from the plane of the sky ($\alpha$) for Abell 3528 ({\em left}),
1750 ({\em middle}) and 3395 ({\em right}). The dashed lines
are the limiting cases for bound systems, thus all orbit solutions to
the right are unbound, while those to the left are bound.  The
measured relative velocity from our data is indicated by the solid
vertical lines and its 68\% confidence region is shown with the
cross-hatching.}
\label{fig:orbit}
\end{figure*}

Since only A3395 shows significantly higher gas temperatures between
the sub-clusters, which we interpret as being due to the merger of
these sub-clusters, we tested whether or not the systems were bound by
examining the overall dynamics of these clusters. A bound system must
necessarily fulfill the simple Newtonian energy consideration:
\begin{equation}
V_r^2R_p \leq 2GM\sin^2\alpha \cos\alpha,
\end{equation}
where $V_r$ is the observed relative radial velocity, $R_p$ is the
projected separation of the two sub-clusters, M is the sum of the
masses of the two sub-clumps (i.e. the entire system) and $\alpha$ is
the projection angle from the plane of the sky. Plots for all three
clusters are presented in Figure~\ref{fig:orbit}.

The dashed line separates the bound configurations (to the left) from
the unbound configurations. We included our measurement of the
relative velocity as a solid vertical line with a one sigma confidence
region marked by cross-hashings.  From Figure~\ref{fig:orbit} it is
clear that every reasonable configuration for both A3528 and A3395
indicates that the systems are bound.


\placefigure{fig:orbit}

For A1750 the dynamical analysis is less conclusive. This is in
contrast to similar previous work on this cluster by Beers et
al. (1991).  While our mass for this cluster is only slightly smaller
than their ``doubled'' mass simulation (6.1 vs 6.8 $\rm \times
10^{14}\ M_\odot$), the larger $R_p$, implied by our cosmology, and
the increase in $V_r$, from our improved velocity sample decrease the
likelihood that the system is bound. However, we note that A1750 has
two other components to the system (Beers et al. 1991, Einasto et
al. 1997, Jones \& Forman 1999) whose mass had not been included in
our simple model. The inclusion of these masses would certainly
strongly increase the likelihood that a purely dynamical analysis
would find the system to be bound.

\section{Summary}\label{sec:summary}
Using observations from {\em ROSAT} and {\em ASCA}, we developed
temperature and density distributions for the hot X-ray emitting gas
in three binary galaxy clusters: A3528, 1750 and 3395.

We find that the values for $\beta$, $R_c$ and the luminosities, as
well as the masses derived from these parameters, are typical of
single clusters. For A3528 and A1750 this is not particularly
surprising, given that the bulk of the photons which determine the fit
are derived from the core regions which in general are much less
distorted than the outer isophotes. However, it is interesting that
A3395 which is clearly strongly distorted, still maintains an
intensity profile typical of relaxed unperturbed clusters.

Using new velocity data, supplemented with measurements from the
literature for each cluster, we generated estimates of the mean
velocities for the sub-clusters as well as for the overall cluster,
estimates of the dispersions (the ``scale'' of the distributions) and
velocity histograms.  Two of the clusters (A3528 and A3395) show no
significant differences in the velocities of the subclusters and thus
the in-fall appears to be nearly in the plane of the sky. For A1750 the
velocity difference of the subclusters is $\rm 1335\ km\ s^{-1}$,
suggesting that the merger lies more along the line of sight. The masses
estimated using the virial method are consistent with the X-ray
derived masses.

We also produced an analysis of their orbital dynamics from simple
Newtonian energy considerations and find that A3528 and A3395 are
nearly certainly bound to each other. The results for A1750 are less
conclusive, however as we note there are several nearby additional mass
components that are not included in our estimates of the total mass
but which are likely part of the same structure and would strongly
increase the probability that the system is bound.

The temperature results indicate some heating of the intra-cluster
gas, consistent with the level of disruption of the surface brightness
distributions. The most prominent feature is the heating of the gas
located between the subclusters of each binary. In addition, the
absence of cool gas in the cores of three of the subclusters (both
parts of A3395 and the southwest element of A1750) may have resulted
from the disruption of the cooling flow by the merger.

Our observations suggest that A3528 is in the early stages of a
merger, where the gas in both sub-clusters is only just beginning to
interact. Both cores show evidence of cool gas, and the intensity
contours are still azimuthally symmetric. The temperature of the gas
located between the sub-clusters is marginally hotter ($\sim 15$\%)
than the overall average for the cluster.

A1750 is slightly further along in the merging process. One of the
sub-clusters retains cool gas in the core, but the other most likely
does not. There is some elongation of the intensity isophotes and the
gas between the cores shows significant heating to $\sim 30$\% above
the overall average.

Finally, A3395 is nearly at first core passage. Neither core shows
evidence for cool gas, the intensity isophotes are highly disrupted
and the temperature distribution clearly shows that the gas between
the two sub-clusters has been heated $\sim 30$\% above the overall
average.

Taken together these three binary clusters present a sequence of views
of a typical merger event in a galaxy cluster.  More detailed
spectroscopic studies with {\em CHANDRA} and {\em XMM} will provide
details of the merging process and its effects on both the
intra-cluster medium and the resident galaxies.

\begin{deluxetable}{cccccccccc}
\tablecolumns{10}
\tablecaption{Observational Data\label{tab:obsdat}}
\tablehead{&&&\multicolumn{3}{c}{ROSAT}&&\multicolumn{3}{c}{ASCA}\nl
\cline{4-6}\cline{8-10}
\multirow{2}{*}{Cluster}&$\alpha$&$\delta$&\multirow{2}{*}{Sequence
\# }&\multirow{2}{*}{Date}&Live Time&&\multirow{2}{*}{Sequence
\# }&\multirow{2}{*}{Date}&On Time\nl
&(J2000)&(J2000)&&&(ksec)&&&&(ksec)}
\startdata
A3528&12:54:29&-29:07:00&300093&7/16/96&15.0&
&84057000& 1/15/96&18.9\nl
A1750&13:31:00&-1:47:18&800553&7/01/93 &12.7&
&81010000& 1/27/94&33.5\nl
A3395&6:27:03&-54:08:48&800079&7/16/91&2.6 &
&82033000& 2/14/95&31.1
\enddata
\end{deluxetable}


\begin{deluxetable}{cccccccc}
\tablecolumns{8}
\tablecaption{Cluster Properties\label{tab:struc+T}}
\tablehead{\multirow{2}{*}{Cluster}&\multirow{2}{*}{$z$}&$L$\tablenotemark{a}&$T_{ASCA}$&\multirow{2}{*}{$\beta$}&$r_c$&$n_e(0)$&$~~~\left\langle\frac{1}{r}\right\rangle^{-1}$\nl
&&($\rm 10^{44}\ ergs\ s^{-1}$)&(keV)&&(Mpc)&$(\rm 10^{-3}\ cm^{-3})$&(Mpc)}
\startdata
A3528 SE
&\multirow{2}{*}{0.0545}
&1.78
&\multirow{2}{*}{$4.7\pm 0.3$}
&$0.55^{+0.02}_{-0.01}$
&$0.14^{+0.02}_{-0.02}$
&3.12
&0.35(0.39)\tablenotemark{b}\nl

A3528 NW
&
&1.26
&
&$0.59^{+0.04}_{-0.04}$
&$0.14^{+0.03}_{-0.02}$
&2.86
&0.22
\nl
\nl

A1750 NE
&\multirow{2}{*}{0.0855}
&1.08
&\multirow{2}{*}{$3.6\pm 0.2$}&
$0.52^{+0.04}_{-0.03}$&
$0.17^{+0.04}_{-0.02}$
&1.75
&0.20
\nl

A1750 SW
&
&1.48
&
&$0.65^{+0.04}_{-0.04}$
&$0.23^{+0.04}_{-0.02}$
&2.02
&0.22
\nl
\nl

A3395 NE
&\multirow{2}{*}{0.0506}
&1.29
&\multirow{2}{*}{$4.5\pm0.2$}
&$0.66^{+0.26}_{-0.13}$
&$0.32^{+0.18}_{-0.11}$
&1.19
&0.23
\nl

A3395 SW
&
&1.39
&
&$0.60^{+0.23}_{-0.12}$
&$0.28^{+0.18}_{-0.11}$
&1.22
&0.28
\nl

\enddata
\tablenotetext{a}{Luminosities are bolometric values from {\em ROSAT}
data within 1 Mpc of each sub-cluster core.}
\tablenotetext{b}{The parenthetical value excludes the two galaxies
with highly discrepant velocities.}

\end{deluxetable}

\begin{deluxetable}{ccll||ccll}
\tablewidth{6.5in}
\tablecolumns{8}
\tablecaption{Abell 3528 Velocities\label{tab:vel1}}
\tablehead{$\alpha$&$\delta$&\multicolumn{1}{c}{$v$}&\multicolumn{1}{c}{\multirow{2}{*}{Reference}}&
$\alpha$&$\delta$&\multicolumn{1}{c}{$v$}&\multicolumn{1}{c}{\multirow{2}{*}{Reference}}\nl
(J2000)&(J2000)&\multicolumn{1}{c}{$\rm (km\ s^{-1})$}&&(J2000)&(J2000)&\multicolumn{1}{c}{$\rm (km\ s^{-1})$}&}	
\startdata
12:52:37.8 & -28:54:47 & $16052\pm 50$ &1&12:54:35.7 & -29:06:38 & $16254\pm 60$ &1\nl
12:53:15.9 & -29:19:04 & $16403\pm 50$ &1&12:54:38.4 & -28:58:56 & $17101\pm 50$ &1\nl
12:53:26.1 & -28:54:11 & $16336\pm 61$ &1&12:54:39.8 & -28:56:32 & $15759\pm 51$ &1\nl
12:53:31.2 & -29:27:18 & $15201\pm 61$ &1&12:54:39.8 & -29:27:33 & $15045\pm 59$ &1\nl
12:53:40.6 & -28:48:58 & $16633\pm132$ &1&12:54:40.5 & -29:01:49 & $15827\pm 50$ &2,3\nl
12:53:45.6 & -29:02:53 & $14872\pm 48$ &2&12:54:41.0 & -28:53:00 & $16432\pm 50$ &1\nl
12:53:52.0 & -28:36:15 & $14791\pm 50$ &1&12:54:41.0 & -29:13:37 & $16783\pm123$ &1\nl
12:53:54.6 & -28:34:04 & $14671\pm 50$ &1&12:54:42.5 & -29:01:00 & $22737\pm 50$ &1\nl
12:53:58.2 & -29:31:09 & $17631\pm 52$ &1&12:54:43.1 & -28:52:16 & $16237\pm 50$ &1\nl
12:54:00.8 & -28:44:07 & $16680\pm 50$ &1&12:54:48.6 & -29:12:00 & $15824\pm 79$ &1\nl
12:54:01.6 & -28:30:00 & $16528\pm 50$ &1&12:54:49.2 & -29:09:54 & $17260\pm 70$ &1\nl
12:54:03.9 & -29:04:22 & $17303\pm 50$ &1&12:54:50.7 & -28:41:47 & $14560\pm 50$ &1\nl
12:54:10.6 & -29:10:41 & $17177\pm 50$ &1&12:54:52.2 & -29:16:16 & $14486\pm 50$ &1\nl
12:54:11.3 & -29:01:46 & $16622\pm 50$ &1&12:54:52.4 & -28:43:17 & $16986\pm 73$ &1\nl
12:54:12.1 & -29:09:22 & $15894\pm 50$ &1&12:54:53.2 & -28:31:58 & $15426\pm 50$ &1\nl
12:54:14.3 & -28:59:16 & $16784\pm 50$ &1&12:54:53.9 & -29:00:46 & $16174\pm 70$ &1\nl
12:54:15.6 & -29:00:19 & $15437\pm 86$ &1&12:54:55.3 & -29:30:11 & $16329\pm137$ &1\nl
12:54:16.3 & -28:45:23 & $15548\pm 50$ &1&12:54:56.2 & -28:55:40 & $15926\pm 50$ &1\nl
12:54:16.9 & -29:01:16 & $15119\pm 50$ &1&12:54:56.4 & -29:06:28 & $13816\pm100$ &1\nl
12:54:17.6 & -29:00:47 & $16186\pm113$ &1&12:55:04.3 & -29:15:56 & $17007\pm 57$ &2\nl
12:54:18.9 & -29:18:11 & $16083\pm 54$ &2&12:55:18.6 & -29:12:04 & $16866\pm 68$ &1\nl
12:54:20.5 & -29:04:10 & $16287\pm 50$ &1&12:55:19.7 & -29:17:20 & $16476\pm 54$ &2\nl
12:54:21.5 & -29:13:23 & $16834\pm 50$ &1&12:55:21.0 & -28:57:01 & $15834\pm 65$ &1\nl
12:54:22.2 & -29:00:45 & $16219\pm 50$ &1&12:55:27.6 & -28:47:50 & $17487\pm 51$ &1\nl
12:54:22.9 & -29:04:17 & $17391\pm 71$ &1&12:55:29.6 & -28:45:35 & $17367\pm109$ &1\nl
12:54:23.3 & -29:01:05 & $16353\pm 63$ &3&12:55:33.0 & -28:48:52 & $17740\pm 69$ &1\nl
12:54:23.5 & -29:10:02 & $16530\pm 50$ &1&12:55:40.2 & -28:34:33 & $17073\pm 50$ &1\nl
12:54:24.7 & -28:59:36 & $16914\pm 50$ &1&12:55:43.6 & -29:24:41 & $18479\pm 94$ &1\nl
12:54:25.3 & -28:58:24 & $14664\pm 50$ &1&12:55:44.8 & -29:01:11 & $16810\pm 50$ &1\nl
12:54:26.7 & -28:57:20 & $15720\pm 50$ &1&12:55:50.8 & -28:45:16 & $16298\pm 50$ &1\nl
12:54:28.1 & -28:57:42 & $14122\pm 50$ &1&12:55:57.5 & -29:17:58 & $16757\pm 57$ &1\nl
12:54:28.1 & -29:00:42 & $13413\pm260$ &1&12:56:13.0 & -29:13:44 & $16825\pm118$ &1\nl
12:54:28.8 & -29:17:31 & $14971\pm 63$ &1&12:56:20.9 & -28:59:14 & $17885\pm 54$ &1\nl
12:54:31.5 & -28:55:18 & $16790\pm 50$ &1&12:56:24.9 & -29:12:59 & $16963\pm 50$ &1\nl
12:54:35.3 & -29:04:41 & $16217\pm 55$ &1&12:56:27.6 & -29:08:54 & $17825\pm 79$ &1\nl
12:54:35.4 & -29:00:39 & $14713\pm 50$ &1&12:56:27.9 & -29:08:55 & $17719\pm 50$ &1\nl
\enddata
\tablerefs{
(1)Quintana et al. 2000; (2)Katgert et al. 1998 ; (3)Quintana et al. 1995}
\end{deluxetable}

\begin{deluxetable}{ccll||ccll}
\tablewidth{6.5in}
\tablecolumns{8}
\tablecaption{Abell 1750 Velocities\label{tab:vel2}}
\tablehead{$\alpha$&$\delta$&\multicolumn{1}{c}{$v$}&\multicolumn{1}{c}{\multirow{2}{*}{Reference}}&
$\alpha$&$\delta$&\multicolumn{1}{c}{$v$}&\multicolumn{1}{c}{\multirow{2}{*}{Reference}}\nl
(J2000)&(J2000)&\multicolumn{1}{c}{$\rm (km\ s^{-1})$}&&(J2000)&(J2000)&\multicolumn{1}{c}{$\rm (km\ s^{-1})$}&}	
\startdata
13:30:15.2 & -1:31:09 & $27131\pm28$& 1  &13:30:57.4 & -1:44:31 & $22939\pm20$& 1,2\nl
13:30:18.4 & -2:00:48 & $26145\pm23$& 1,2&13:30:58.2 & -1:51:13 & $25413\pm50$& 1  \nl
13:30:18.7 & -1:32:44 & $24826\pm22$& 1  &13:30:59.8 & -1:43:36 & $24780\pm40$& 1,2\nl
13:30:32.3 & -1:47:36 & $26297\pm28$& 1  &13:31:01.2 & -1:50:52 & $26012\pm30$& 1,2\nl
13:30:32.9 & -1:39:35 & $25171\pm35$& 1  &13:31:04.7 & -1:38:04 & $24665\pm33$& 1  \nl
13:30:34.4 & -1:47:45 & $25034\pm34$& 1  &13:31:06.1 & -1:40:48 & $25184\pm30$& 1,2\nl
13:30:34.5 & -1:48:27 &$26354\pm100$& 2  &13:31:06.4 & -1:44:14 & $24400\pm50$& 2  \nl
13:30:35.1 & -1:49:05 & $24543\pm36$& 1  &13:31:09.7 & -1:44:45 & $25229\pm50$& 2  \nl
13:30:36.8 & -1:53:04 & $25951\pm23$& 1,2&13:31:09.9 & -1:41:09 & $24510\pm20$& 1,2\nl
13:30:38.2 & -1:50:36 & $25598\pm24$& 1,2&13:31:10.4 & -1:38:03 & $25880\pm37$& 1  \nl
13:30:48.9 & -1:53:56 & $27585\pm46$& 1,2&13:31:10.8 & -1:43:49 & $25019\pm41$& 1,2\nl
13:30:41.4 & -1:41:14 & $26165\pm44$& 1  &13:31:11.0 & -1:43:41 & $25058\pm26$& 1,2\nl
13:30:42.5 & -1:51:24 & $25247\pm26$& 1,2&13:31:11.2 & -1:43:35 & $25381\pm24$& 1,2\nl
13:30:44.4 & -1:53:19 & $26408\pm23$& 1,2&13:31:13.3 & -1:58:42 & $26903\pm38$& 1  \nl
13:30:44.5 & -1:59:29 & $24809\pm30$& 1  &13:31:13.9 & -1:48:58 & $25740\pm38$& 1  \nl
13:30:45.5 & -1:48:43 & $24051\pm57$& 2  &13:31:14.7 & -1:30:17 & $25461\pm37$& 1  \nl
13:30:46.1 & -1:52:28 & $26434\pm31$& 1,2&13:31:16.2 & -1:59:35 & $27684\pm43$& 1  \nl
13:30:48.1 & -1:52:02 & $26231\pm32$& 1,2&13:31:19.3 & -1:42:19 & $24339\pm50$& 2  \nl
13:30:48.1 & -1:52:12 & $25118\pm25$& 1,2&13:31:19.5 & -1:45:39 & $24522\pm27$& 1,2\nl
13:30:48.5 & -1:37:48 & $25100\pm42$& 1  &13:31:22.2 & -1:43:14 & $24933\pm50$& 2  \nl
13:30:49.0 & -1:52:52 & $27926\pm50$& 2  &13:31:23.1 & -2:01:37 & $26731\pm45$& 1  \nl
13:30:49.1 & -1:53:56 & $27361\pm50$& 2  &13:31:23.4 & -1:43:35 & $22724\pm23$& 1,2\nl
13:30:49.8 & -1:40:40 & $24328\pm20$& 1  &13:31:23.9 & -1:53:35 & $26343\pm37$& 1  \nl
13:30:50.1 & -1:47:55 & $26260\pm30$& 1,2&13:31:27.9 & -1:43:50 & $25260\pm41$& 2  \nl
13:30:50.4 & -1:51:27 & $26086\pm24$& 1,2&13:31:30.0 & -1:39:25 & $25259\pm24$& 1,2\nl
13:30:50.5 & -1:51:44 & $26472\pm29$& 1,2&13:31:36.6 & -1:40:08 & $24623\pm43$& 2  \nl
13:30:50.8 & -1:51:37 & $26088\pm26$& 1,2&13:31:37.7 & -1:52:25 & $26268\pm39$& 1  \nl
13:30:53.0 & -1:37:31 & $24965\pm30$& 1  &13:31:49.5 & -1:58:52 & $26610\pm31$& 1  \nl
13:30:54.2 & -1:52:33 & $26512\pm31$& 1,2&13:31:50.1 & -1:57:38 & $27081\pm30$& 1  \nl
13:30:55.6 & -1:55:13 & $23632\pm43$& 1  &13:31:50.9 & -1:41:15 & $24889\pm35$& 1  \nl
13:30:56.3 & -1:54:08 & $26873\pm50$& 2  &13:31:52.1 & -1:56:07 & $26789\pm28$& 1  \nl
13:30:56.4 & -1:42:20 & $27244\pm50$& 2  &13:31:54.4 & -1:43:44 & $25915\pm21$& 1  \nl  
13:30:56.6 & -1:43:24 & $24704\pm23$& 1,2&13:31:56.7 & -1:49:35 & $22697\pm44$& 1  \nl
13:30:56.7 & -1:53:58 & $25515\pm29$& 1,2&13:32:15.5 & -1:43:13 & $26803\pm22$& 1  \nl
\enddata
\tablerefs{
(1) LCO-SHECTOGRAPH; (2)Beers et al. 1991}
\end{deluxetable}

\begin{deluxetable}{ccll||ccll}
\tablewidth{6.5in}
\tablecolumns{8}
\tablecaption{Abell 3395 Velocities\label{tab:vel3}}
\tablehead{$\alpha$&$\delta$&\multicolumn{1}{c}{$v$}&\multicolumn{1}{c}{\multirow{2}{*}{Reference}}&
$\alpha$&$\delta$&\multicolumn{1}{c}{$v$}&\multicolumn{1}{c}{\multirow{2}{*}{Reference}}\nl
(J2000)&(J2000)&\multicolumn{1}{c}{$\rm (km\ s^{-1})$}&&(J2000)&(J2000)&\multicolumn{1}{c}{$\rm (km\ s^{-1})$}&}	
\startdata
6:24:23.2&-54:29:48&$14478\pm105$&1      & 6:27:23.7&-54:28:23&$15743\pm 70$&1,2    \nl
6:24:26.9&-54:32:55&$15741\pm 71$&1      & 6:27:24.4&-54:26:30&$14440\pm 32$&2      \nl
6:24:30.1&-54:07:36&$15338\pm 69$&1      & 6:27:26.6&-54:25:48&$15099\pm 30$&1,2    \nl
6:24:32.2&-54:10:04&$14012\pm 49$&1      & 6:27:26.3&-54:31:02&$15735\pm 42$&1,4,7  \nl
6:24:35.6&-54:07:47&$14324\pm 48$&1      & 6:27:29.8&-54:27:19&$15024\pm 52$&1,4    \nl
6:24:38.8&-54:14:18&$14410\pm101$&1      & 6:27:30.5&-54:35:48&$17949\pm 39$&1,2    \nl
6:24:48.9&-54:14:48&$13808\pm242$&1,2,3  & 6:27:36.1&-54:24:05&$14675\pm 72$&1      \nl
6:24:60.0&-54:10:31&$17463\pm 30$&1,2    & 6:27:36.3&-54:26:58&$14562\pm 30$&1,2,5,7\nl
6:25:04.0&-54:22:21&$13835\pm 91$&1,2    & 6:27:35.6&-54:40:04&$16106\pm 77$&1      \nl
6:25:04.8&-54:09:41&$17625\pm110$&1      & 6:27:38.2&-54:25:50&$16486\pm118$&1      \nl
6:25:09.3&-54:25:01&$17570\pm 44$&1,2    & 6:27:39.1&-54:22:38&$13754\pm925$&1,4    \nl
6:25:15.3&-54:16:15&$17670\pm276$&1,2    & 6:27:39.4&-54:27:43&$14676\pm 30$&1,2    \nl
6:25:20.1&-54:21:36&$14502\pm108$&1      & 6:27:40.0&-54:26:55&$16222\pm 34$&2      \nl
6:25:22.4&-54:16:54&$17504\pm 57$&1,4    & 6:27:41.0&-54:27:15&$17046\pm 71$&1,4,5  \nl
6:25:31.5&-54:34:41&$15088\pm105$&1      & 6:27:41.3&-54:23:48&$16553\pm131$&1      \nl
6:25:48.3&-54:11:43&$14293\pm 82$&1,4    & 6:27:40.5&-54:36:50&$15676\pm 47$&1      \nl
6:25:49.1&-53:59:36&$13910\pm 89$&1,5    & 6:27:41.9&-54:17:25&$14342\pm 78$&1      \nl
6:25:47.8&-54:39:53&$16401\pm 76$&1      & 6:27:40.9&-54:35:34&$15465\pm170$&1,3    \nl
6:25:51.2&-54:18:22&$16134\pm 88$&1      & 6:27:42.0&-54:32:19&$15245\pm163$&1,2    \nl
6:25:57.0&-54:27:50&$18047\pm 91$&1,3,4  & 6:27:44.3&-54:07:10&$14912\pm163$&1,2    \nl
6:25:58.4&-54:27:44&$18263\pm 35$&2      & 6:27:44.2&-54:16:51&$16158\pm 83$&1      \nl
6:26:01.0&-54:27:40&$18153\pm139$&1      & 6:27:43.8&-54:24:26&$13712\pm 33$&1,2    \nl
6:26:08.7&-54:22:04&$14777\pm 92$&1      & 6:27:43.7&-54:27:19&$16248\pm 73$&1,4,5  \nl
6:26:07.9&-54:40:30&$15499\pm 79$&1,2,6  & 6:27:44.6&-54:26:45&$13265\pm 84$&1,2,5  \nl
6:26:10.3&-54:27:24&$15670\pm 43$&1,2    & 6:27:47.0&-54:25:10&$16305\pm201$&1,3,4  \nl
6:26:10.4&-54:32:26&$16418\pm 56$&1,2    & 6:27:47.6&-54:25:28&$14872\pm 52$&1      \nl
6:26:11.5&-54:17:14&$14960\pm 33$&4      & 6:27:49.1&-54:04:03&$17986\pm 41$&2      \nl
6:26:12.0&-54:17:08&$15224\pm 42$&2      & 6:27:48.6&-54:32:19&$13076\pm 55$&4      \nl
6:26:12.1&-54:23:40&$14230\pm106$&1      & 6:27:50.9&-54:09:05&$14586\pm 69$&1      \nl
6:26:11.4&-54:40:16&$17627\pm 98$&1      & 6:27:51.1&-54:09:07&$14926\pm223$&1,2    \nl
6:26:13.7&-54:24:19&$17728\pm 30$&1,2    & 6:27:50.7&-54:28:22&$15535\pm 55$&1,2    \nl
6:26:15.2&-54:19:37&$14659\pm 75$&1      & 6:27:52.5&-54:30:04&$14322\pm311$&1,4    \nl
6:26:16.8&-54:20:17&$14911\pm 31$&1,2    & 6:27:52.1&-54:36:52&$16616\pm104$&1      \nl
6:26:17.3&-54:32:26&$15887\pm 81$&1,2    & 6:27:54.5&-54:31:24&$16152\pm166$&1,2    \nl
6:26:20.7&-54:15:32&$14543\pm112$&1,2,3  & 6:27:56.3&-54:26:02&$13983\pm 66$&4      \nl
6:26:24.1&-54:20:38&$14755\pm 77$&1,2    & 6:27:58.5&-54:32:50&$16143\pm188$&1      \nl
6:26:27.4&-54:28:35&$14797\pm 98$&1      & 6:27:59.3&-54:29:44&$15097\pm104$&1,2    \nl
6:26:30.5&-54:17:57&$13248\pm 81$&1      & 6:28:01.4&-54:35:09&$15360\pm 94$&1      \nl
6:26:31.5&-54:36:55&$14172\pm 50$&2      & 6:28:02.0&-54:34:14&$14153\pm 89$&1,2    \nl
6:26:32.4&-54:34:01&$16305\pm 30$&2      & 6:28:05.8&-54:24:12&$15033\pm 35$&1      \nl
6:26:34.3&-54:10:00&$13544\pm 78$&1,2    & 6:28:08.0&-54:34:18&$15530\pm 76$&1,2    \nl
6:26:34.7&-54:09:50&$14621\pm 35$&2      & 6:28:13.2&-54:15:35&$16437\pm 39$&1,2    \nl
6:26:34.5&-54:21:25&$15264\pm202$&1,3,4  & 6:28:12.5&-54:29:47&$16163\pm 48$&2      \nl
6:26:34.6&-54:25:56&$15548\pm 69$&1      & 6:28:15.2&-54:35:23&$15793\pm134$&1,2,7  \nl
6:26:35.8&-54:29:31&$15853\pm 51$&1,3,4  & 6:28:16.6&-54:31:39&$15909\pm 74$&1,2,7  \nl
6:26:37.5&-54:26:02&$14542\pm150$&1,4    & 6:28:18.7&-54:15:09&$15686\pm 99$&1,3,4  \nl
6:26:40.1&-54:12:03&$15623\pm 81$&1      & 6:28:18.9&-54:21:36&$14548\pm 72$&1      \nl
6:26:40.2&-54:11:23&$17655\pm115$&1      & 6:28:20.3&-54:36:38&$14983\pm153$&1,2    \nl
6:26:40.0&-54:20:02&$14139\pm 30$&1,2    & 6:28:21.2&-54:25:25&$15451\pm238$&1,2    \nl
6:26:40.4&-54:32:48&$18213\pm 49$&2      & 6:28:23.1&-54:20:00&$14591\pm102$&1      \nl
6:26:43.4&-54:19:14&$14675\pm896$&1,2    & 6:28:23.2&-54:35:55&$15432\pm 47$&1,2    \nl
6:26:43.5&-54:26:16&$15500\pm116$&1      & 6:28:29.1&-54:22:48&$14685\pm272$&1,2,3  \nl
6:26:47.9&-54:24:06&$13626\pm 30$&1,2,3  & 6:28:30.0&-54:16:23&$15412\pm 62$&1,4    \nl
6:26:49.5&-54:32:34&$15622\pm 58$&1,2,3  & 6:28:34.6&-54:28:05&$15564\pm142$&1,2    \nl
6:26:51.0&-54:32:48&$15003\pm 40$&1,2    & 6:28:46.8&-54:23:25&$14313\pm 63$&1      \nl
6:26:52.0&-54:28:52&$15503\pm 64$&1,2    & 6:28:49.1&-54:28:16&$13447\pm 40$&1      \nl
6:26:53.6&-54:35:07&$15445\pm 95$&1      & 6:28:57.2&-54:17:17&$15847\pm 30$&1,2    \nl
6:26:54.9&-54:19:14&$14076\pm101$&1      & 6:29:00.2&-54:12:41&$14836\pm163$&1,2    \nl
6:26:57.8&-54:33:31&$15877\pm 31$&1,2    & 6:28:59.7&-54:22:12&$15088\pm158$&1,2    \nl
6:26:58.7&-54:28:09&$15432\pm 92$&1      & 6:28:59.8&-54:23:03&$14561\pm 31$&2      \nl
6:27:04.7&-54:25:47&$15095\pm146$&1      & 6:29:03.0&-54:23:12&$14977\pm 94$&1,2,5,7\nl
6:27:06.5&-54:22:15&$15026\pm180$&1,2    & 6:29:09.9&-54:26:00&$15003\pm186$&1,2    \nl
6:27:06.1&-54:33:26&$15036\pm159$&1,2    & 6:29:27.0&-54:27:03&$14678\pm 39$&2      \nl
6:27:07.0&-54:26:40&$16212\pm113$&1      & 6:29:33.4&-54:15:26&$13311\pm 94$&1,4    \nl
6:27:08.8&-54:05:31&$15300\pm 34$&1,2,3,4& 6:29:37.2&-54:42:46&$13856\pm 71$&4      \nl
6:27:08.8&-54:22:25&$16412\pm 39$&1,2,3  & 6:29:44.7&-54:36:25&$14510\pm 48$&1,2    \nl
6:27:09.0&-54:26:45&$14930\pm 77$&1      & 6:29:47.1&-54:48:18&$14912\pm 69$&1,2,3  \nl
6:27:10.4&-54:20:08&$15537\pm113$&1      & 6:29:47.6&-54:47:48&$15110\pm 73$&2,3    \nl
6:27:15.0&-54:00:03&$15362\pm 88$&1      & 6:30:06.6&-54:35:13&$14506\pm 50$&1      \nl
6:27:14.6&-54:08:34&$16238\pm 69$&1,4    & 6:30:09.9&-54:17:22&$15243\pm 49$&1      \nl
6:27:14.8&-54:25:34&$14849\pm 72$&1,2    & 6:30:15.6&-54:20:56&$16162\pm150$&1      \nl
6:27:16.1&-54:23:00&$16624\pm 32$&1,2    & 6:30:24.3&-54:41:02&$14686\pm 81$&1      \nl
6:27:16.4&-54:32:07&$15808\pm 31$&1,2    & 6:30:25.4&-54:26:47&$13911\pm181$&1      \nl
6:27:19.4&-54:06:58&$17444\pm 36$&1,4    & 6:30:25.6&-54:45:45&$14970\pm 43$&1,2    \nl
6:27:18.9&-54:24:06&$15609\pm123$&4      & 6:30:29.3&-54:33:22&$15377\pm123$&1      \nl
6:27:19.2&-54:25:15&$13657\pm 40$&1,4    & 6:30:28.9&-54:41:06&$14959\pm 67$&1      \nl
6:27:19.7&-54:28:31&$14766\pm262$&1,4    & 6:30:36.3&-54:44:07&$14312\pm106$&1      \nl
6:27:19.6&-54:37:00&$14560\pm 80$&4      & 6:30:51.2&-54:26:00&$13432\pm 85$&1      \nl
6:27:21.0&-54:26:20&$16430\pm 30$&1,2,4  &          &         &&\nl
\enddata			
\tablerefs{
(1) Teague et al. 1990 with 82 $\rm km\ s^{-1}$ offset; (2)
LCO-SHECTOGRAPH; (3) Hopp and Materne 1985; (4) CTIO-ARGUS with 50
$\rm km\ s^{-1}$ offset; (5) Materne et al. 1982; (6) West and
Frandsen 1981; (7) Vidal 1975 
}
\end{deluxetable}

\begin{deluxetable}{cccc}
\tablewidth{4in}
\tablecolumns{4}
\tablecaption{Composite Velocity Results\label{tab:compvel}}
\tablehead{\multirow{2}{*}{Cluster}&$\overline{v}$&$\sigma$&$V_r$\nl
&$(\rm km\ s^{-1})$&$(\rm km\ s^{-1})$&$(\rm km\ s^{-1})$}
\startdata	
Abell 3528&$16346\pm 118$&$985\pm 101$&\nl
SE        &$16040\pm318$&$930\pm 255$($516\pm 151$)\tablenotemark{a}&\multirow{2}{*}{$96^{+379}_{-96}$}\nl
NW	  &$16136\pm207$&$969\pm 198$&\nl
&&&\nl
Abell 1750&$25632\pm 132$&$1050\pm 96$&\nl
NE        &$24815\pm 153$&$676\pm 310$&\multirow{2}{*}{$1335\pm 211$}\nl
SW        &$26150\pm 145$&$906\pm 236$&\nl
&&&\nl
Abell 3395&$15177\pm118$&$1069\pm 81$&\nl
NE all    &$15325\pm167$&$949\pm 103$&\multirow{2}{*}{$102^{+214}_{-102}$}\nl
SW all    &$15427\pm135$&$713\pm 185$&\nl
NE clip   &$15244\pm165$&$1036\pm 107$&\multirow{2}{*}{$192^{+236}_{-192}$}\nl
SW clip   &$15436\pm169$&$939\pm 311$&\nl
\enddata
\tablenotetext{a}{The parenthetical value excludes the two galaxies
with highly discrepant velocities.} 
\end{deluxetable}

\begin{deluxetable}{cccccccc}
\tablecolumns{8}
\tablecaption{Mass\tablenotemark{a}\  Determinations\label{tab:masses}}
\tablehead{\multirow{2}{*}{Cluster}&\multicolumn{2}{c}{$M_{gas}$}&\multicolumn{2}{c}{$M_{total}$}&$M_{isothermal}$&$M_{scaled}$&$M_{virial}$\nl
\cline{2-3}\cline{4-5}
&0.5 $Mpc$&1.0 $Mpc$&0.5 $Mpc$&1.0 $Mpc$&1.0 $Mpc$&1.0 $Mpc$&$Total$} 
\startdata
A3528 SE&$0.09^{+.01}_{-.01}$&$0.27^{+.02}_{-.02}$&$1.7^{+.2}_{-.1}$
&$3.6^{+.3}_{-.3}$&$2.8^{+.1}_{-.2}$&$4.4^{+.3}_{-.3}$
&$6.6^{+4.1}_{-3.1}(2.27^{+1.5}_{-1.2})$\tablenotemark{b}\nl

A3528 NW&$0.08^{+.01}_{-.01}$&$0.22^{+.02}_{-.01}$&$1.9^{+.2}_{-.2}$
&$4.0^{+.3}_{-.4}$&$3.0^{+.1}_{-.2}$&$4.3^{+.4}_{-.4}$
&$4.5^{+2.1}_{-1.6}$\nl\nl

A1750 NE&$0.07^{+.01}_{-.01}$&$0.23^{+.02}_{-.02}$&$1.3^{+.1}_{-.1}$
&$2.7^{+.3}_{-.3}$&$2.2^{+.1}_{-.2}$&$3.7^{+.4}_{-.4}$
&$2.0^{+2.3}_{-1.4}$\nl

A1750 SW&$0.09^{+.01}_{-.01}$&$0.26^{+.02}_{-.02}$&$1.5^{+.2}_{-.1}$
&$3.4^{+.3}_{-.4}$&$2.7^{+.1}_{-.2}$&$3.6^{+.4}_{-.3}$
&$4.0^{+2.3}_{-1.1}$\nl\nl

A3395 NE&$0.08^{+.01}_{-.01}$&$0.25^{+.02}_{-.02}$&$1.6^{+.1}_{-.2}$
&$3.8^{+.3}_{-.3}$&$3.0^{+.1}_{-.1}$&$4.2^{+.4}_{-.3}$
&$4.5^{+1.1}_{-0.9}$\nl

A3395 SW&$0.08^{+.01}_{-.01}$&$0.26^{+.02}_{-.02}$&$1.5^{+.1}_{-.2}$
&$3.4^{+.3}_{-.3}$&$2.6^{+.1}_{-.2}$&$4.2^{+.5}_{-.3}$
&$3.1^{+1.8}_{-1.4}$
\enddata
\tablenotetext{a}{Masses are all in units of $10^{14}\ M_\odot$.}
\tablenotetext{b}{The parenthetical value excludes the two galaxies
with highly discrepant velocities.}
\end{deluxetable}

\acknowledgments 

The authors would like to thank Dan Fabricant for helpful
comments. RHD, WF, CJ acknowledge support from the Smithsonian
Institute and NASA contract NAS8-39073. HQ acknowledges partial
support from FONDECYT grant 8970009 and the award of a Presidential
Chair in Science (Chile).

\newpage
\begin{figure*}[t]
\centerline{
}
\caption{{\em Top left:} Continuous temperature map for A3528 from
{\em ASCA} GIS data with intensity contours and region identifiers
overlaid. {\em Bottom left:} Fit temperatures with 90\% confidence
error bars for the regions defined on the continuous temperature
map. The global cluster temperature is show as a dotted line.{\em Top
right:} Galaxies with redshifts are overlain on the {\em ROSAT}
intensity contours. Galaxies with velocities greater than the overall
mean cluster velocity are shown in red and those with smaller
velocities are in blue. The field-of-view for the top right has been
expanded to include all of the galaxies listed in Table~\ref{tab:vel2}
and shown in the lower right. The outline of the top left frame is
shown as a dashed line. Circles denoting a radius of a half Mpc are
shown centered on each sub-cluster core and are color coded to match
the sub-cluster histograms shown in the bottom right. {\em Bottom
right:} Velocity histogram of the galaxies shown in the top
right. Bins are 300 $\rm km\ s^{-1}$ wide and the overall cluster
velocity is indicated by a dashed line.}
\label{fig:3528}
\end{figure*}

\newpage
\begin{figure*}[t]
\centerline{
}
\caption{Same as for Figure~\ref{fig:3528} for Abell 1750.  The
field-of-view for the top right has been expanded to include all of
the galaxies listed in Table~\ref{tab:vel2} and shown in the lower
right; the outline of the top left frame is shown as a dashed
line. The half Mpc circles in the top right are color coded to match
the velocity histograms of the sub-cluster cores shown in the bottom
right.}
\label{fig:1750}
\end{figure*}

\newpage
\begin{figure*}[t]
\centerline{
}
\caption{Same as for Figure~\ref{fig:3528} for Abell 3395. The
field-of-view for the top right has been expanded to include all of
the galaxies listed in Table~\ref{tab:vel3} and shown in the lower
right; the outline of the top left frame is shown as a dashed
line. The half Mpc circles in the top right are color coded to match
the velocity histograms of the sub-cluster cores shown in the bottom
right. The galaxies in the high velocity bump in the bottom right are
shown as empty circles in the top right panel.}
\label{fig:3395}
\end{figure*}

\newpage


\begin{references}
\reference{}Abell, G.O., Corwin, H.G. \& Olowin, R.P. 1989, \apjs, 70, 1.
\reference{}Arnaud, K. 1993, {\em ASCA} Newsletter, No. 1 (NASA/GSFC).
\reference{}Arnaud, M. Neumann, D.M., Aghanim, N., Gastaud, R.,
Majerowicz, S., \& Hughes, J.P. 2001, A\&A, 365, L80.
\reference{}Beers, T.C., Geller, M.J. \& Huchra, J.P. 1982, ApJ, 257, 23.
\reference{}Beers, T.C., Flynn, K., \& Gebhardt, K. 1990, AJ, 100, 32.
\reference{}Beers, T., Forman, W., Huchra, J., Jones, C. \& Gebhardt,
K. 1991, AJ, 102, 1581
\reference{}Bird, C.~M. 1994, AJ, 107, 1637.
\reference{}Cavaliere, A. \& Fusco-Femiano, R. 1976, A\&A, 49, 137.
\reference{}Churazov, E., Gilfanov, M., Forman, W. \& Jones, C. 1996,
ApJ, 471, 673
\reference{}Churazov, E., Gilfanov, M., Forman, W. \& Jones, C. 1999,
ApJ, 520, 105
\reference{}David, L.P., Arnaud, K.A., Forman, W. \& Jones, C. 1990,
ApJ, 356, 32
\reference{}Donnelly, R. Hank, Markevitch, M., Forman, W., Jones, C.,
Churazov, E. \& Gilfanov, M. 1999, ApJ, 513, 690.
\reference{}Donnelly, R. Hank, Markevitch, M., Forman, W., Jones, C.,
David, L.P., Churazov, E. \& Gilfanov, M. 1998, ApJ, 500, 138.
\reference{}Dressler, A. \& Shectman, S.~A. 1988, AJ, 985.
\reference{}Einasto, M., Tago, E., Jaaniste, J., Einasto, J., \&
Andernach, H. 1997, A\&AS, 123, 119.
\reference{9}Evrard, A.~E. 1990a, ApJ, 363, 349.
\reference{10}Evrard, A.~E. 1990b, in Clusters of Galaxies, eds. Oegerle,
Fitchett and Danly (Cambridge: Cambridge University Press), 287.
\reference{}Evrard, A.E., Metzler, C.A. \& Navarro, J.F. 1996, ApJ,
469, 490.
\reference{}Fabricant, D.G., Bautz, M.W. \& McClintock, J.E. 1994,
AJ, 107, 8.
\reference{}Forman, C. \& Jones, C. 1990, in Clusters of Galaxies,
ed. W. Oergerle, M. Fitchett, \& L. Danly (Cambridge: Cambridge Univ.
Press), 257.
\reference{12}Forman, W., Bechtold, J., Blair, W., Giacconni, R., Van
Speybroeck, L., \& Jones, C. 1981, ApJL, 243, L133.
\reference{14}Geller, M.J., \& Beers, T.C. 1982, PASP, 94, 421.
\reference{}Henricksen, M., Donnelly, R. Hank \& Davis, D. 2000, ApJ,
529, 692.
\reference{}Henriksen, M. \& Markevitch, M. 1996, ApJL, 466, L79.
\reference{}Henriksen, M. \& Jones, C. 1996, ApJ, 465, 666.
\reference{}Henry, J.P. \& Briel, U.G. 1995, ApJL, 443, L9.
\reference{}Hopp, U. and Materne, J. 1985, A\&A, 148, 359
\reference{}Jones, C. \& Forman, W.R. 1999, ApJ, 511, 65.
\reference{}Jones, C. \& Forman, W.R. 1984, ApJ, 276, 38.
\reference{}Katgert, P., Mazure, A., Perea, J., den Hartog, R., Moles, M., Le Fevre, O., Dubath, P., Focardi, P., Rhee, G., Jones, B., Escalera, E., Biviano, A., Gerbal, D. \& Giuricin, G. 1996, A\&A, 310,8.
\reference{}Katgert, P., Mazure, A., den Hartog, R., Adami, C.,
Biviano, A., and Perea, J. 1998, A\&AS, 129, 399.
\reference{}Maddox, S.J., Efstathiou, G., Sutherland, W.J. and
Loveday, J. 1990, MNRAS, 243, 692.
\reference{}Markevitch, M., Ponman, T. J., Nulsen, P.E.J., Bautz,
M.W., Burke, D.J., David, L.P., Davis, D., Donnelly, R.H., Forman,
W.R., Jones, C., Kaastra, J., Kellogg, E., Kim, D.-W., Kolodziejczak,
J., Mazzotta, P., Pagliaro, A., Patel, S., Van Speybroeck, L.,
Vikhlinin, A., Vrtilek, J., Wise, M. \& Zhao, P. 2000, ApJ, 541, 542.
\reference{}Markevitch, M., Vikhlinin, A., Forman, W.R \& Sarazin,
C.L. 1999, ApJ, 527, 545.
\reference{}Markevitch, M., Forman, W.R., Sarazin, C.L. \& Vikhlinin,
A. 1998, ApJ 503, 77.
\reference{}Markevitch, M., Sarazin, C. \& Irwin, J.A. 1996a, ApJ, 472, L17. 
\reference{}Markevitch, M., Mushotzky, R., Inoue, H., Yamashita, K.,
Furuzawa, A., \& Tawara, Y. 1996b, ApJ, 456, 437
\reference{}Materne, J., Chincarini, G., Tarenghi, M. \& Hopp,
U. 1982, A\&A, 109, 238
\reference{}Mohr, J., Fabricant, D.G., Geller, M.~J., and Evrard, A.E. 1995,
ApJ, 447, 8.
\reference{}Mohr, J.J., Reese, E.D., Ellingson, E., Lewis, A.D. \&
Evrard, A.E. 2000, ApJ,, 544, 109.
\reference{}Norman, M.L. \& Bryan, G.L. 1999, in ``The Radio Galaxy
Messier 87'', ed. R\"{o}ser, H.-J. \& Meisenheimer, K. ( New York,
Springer), p. 106
\reference{}Quintana H., Ram\'{\i}rez A., 1990, AJ, 100, 1424. 
\reference{}Quintana H., Ram\'{\i}rez, A., Melnick, J., Raychaudhury,
S., \& Slezak, E. 1995, AJ, 110, 463.
\reference{}Quintana H., Ram\'{\i}rez, A., \& Way M.J. 1996, AJ, 112, 36.
.\reference{}Quintana H., Carrasco, E.R., \& Reisenegger, A. 2000, AJ,
120, 511.
\reference{}Raychaudury, S., Fabian, A.C., Edge, A.C., Jones, C. \&
Forman, W. 1991, MNRAS, 248, 101.
\reference{}Richstone, Loeb \& Turner 1992, ApJ, 393, 477. 
\reference{}Ricker, P.M. 1998, ApJ , 496, 670.
\reference{}Roettiger, K., Burns, J.O., \& Loken, C.  1996, ApJ, 473,
651.
\reference{}Roettiger, K., Loken, C. \& Burns, J.O. 1997, \apjs, 109, 307.
\reference{}Schindler, S., \& M\"{u}ller, E. 1993, A\&A, 272, 137.
\reference{}Slezak, E.,  Durret, F. \& Gerbal, D. 1994, AJ, 108, 1996.
\reference{}Snowden, S.L. 1994, Cookbook for Analysis Procedures for
{\em ROSAT} XRT/PSPC Observations of Extended Objects and the Diffuse
Background.
\reference{}Snowden, S.L., McCammon, D., Burrows, D.N. \& Mendenthall,
J.A. 1994, ApJ, 424, 714.
\reference{}Struble, M.F. \& Rood, H.J. 1999, \apjs, 125, 35.
\reference{}Takahashi, T., Markevitch, M., Fukuzawa, Y., Ikebe, Y.,
Ishiaki, Y., Kikuchi, K., Makishimi, K. \& Tarawa, Y. 1995, {\em ASCA}
Newsletter, No. 3 (NASA/GSFC).
\reference{}Teague, P.F., Carter, D. \& Gray, P. 1990, \apjs, 72, 715
\reference{}Vidal, N.V. 1975 PASP, 87, 625
\reference{}Vikhlinin, A., Forman, W. \& Jones, C. 1999, ApJ, 525, 47.
\reference{}West and Frandsen 1981, A\&AS, 44, 329
\reference{}White, D.A., Jones, C. \& Forman, W. 1997, MNRAS, 292, 419.
\reference{}White, D.A. 1999 {\it private communication}.
\end{references}
\end{document}